\documentclass[pra,twocolumn,showpacs,superscriptaddress,floatfix, nofootinbib]{revtex4-2}
\usepackage[caption=false]{subfig}
\usepackage{amsmath,graphicx}
\usepackage{amsmath,amssymb,mathrsfs,esint}
\usepackage{multirow}
\usepackage{algorithm}
\usepackage[noend]{algpseudocode}
\usepackage{comment}
\usepackage{tabularx}
\usepackage{subfig}
\usepackage{graphicx}
\usepackage{bm}
\usepackage[normalem]{ulem}
\usepackage[bookmarks=true,
   colorlinks=true,
   linkcolor=blue,
   urlcolor=blue,
   citecolor=blue,
   bookmarks=true,
   hyperindex=true
]{hyperref}

\begin{document}

\title{Continuous dynamical decoupling of optical $^{171}$Yb$^{+}$ qudits with radiofrequency fields}

\author{Ilia V. Zalivako}\email{zalivako.ilya@yandex.ru}
\affiliation{P.N. Lebedev Physical Institute of the Russian Academy of Sciences, Moscow 119991, Russia}
\affiliation{Russian Quantum Center, Skolkovo, Moscow 121205, Russia}

\author{Alexander S. Borisenko}
\affiliation{P.N. Lebedev Physical Institute of the Russian Academy of Sciences, Moscow 119991, Russia}
\affiliation{Russian Quantum Center, Skolkovo, Moscow 121205, Russia}

\author{Ilya A. Semerikov}
\affiliation{P.N. Lebedev Physical Institute of the Russian Academy of Sciences, Moscow 119991, Russia}
\affiliation{Russian Quantum Center, Skolkovo, Moscow 121205, Russia}

\author{Andrey Korolkov}
\affiliation{P.N. Lebedev Physical Institute of the Russian Academy of Sciences, Moscow 119991, Russia}
\affiliation{Russian Quantum Center, Skolkovo, Moscow 121205, Russia}

\author{Pavel L. Sidorov}
\affiliation{P.N. Lebedev Physical Institute of the Russian Academy of Sciences, Moscow 119991, Russia}
\affiliation{Russian Quantum Center, Skolkovo, Moscow 121205, Russia}

\author{Kristina Galstyan}
\affiliation{P.N. Lebedev Physical Institute of the Russian Academy of Sciences, Moscow 119991, Russia}
\affiliation{Russian Quantum Center, Skolkovo, Moscow 121205, Russia}

\author{Nikita V. Semenin}
\affiliation{P.N. Lebedev Physical Institute of the Russian Academy of Sciences, Moscow 119991, Russia}
\affiliation{Russian Quantum Center, Skolkovo, Moscow 121205, Russia}

\author{Vasiliy Smirnov}
\affiliation{P.N. Lebedev Physical Institute of the Russian Academy of Sciences, Moscow 119991, Russia}
\affiliation{Russian Quantum Center, Skolkovo, Moscow 121205, Russia}

\author{Mikhail A.~Aksenov}
\affiliation{P.N. Lebedev Physical Institute of the Russian Academy of Sciences, Moscow 119991, Russia}
\affiliation{Russian Quantum Center, Skolkovo, Moscow 121205, Russia}

\author{Aleksey K.~Fedorov}
\affiliation{P.N. Lebedev Physical Institute of the Russian Academy of Sciences, Moscow 119991, Russia}
\affiliation{Russian Quantum Center, Skolkovo, Moscow 121205, Russia}

\author{Ksenia Yu.~Khabarova}
\affiliation{P.N. Lebedev Physical Institute of the Russian Academy of Sciences, Moscow 119991, Russia}
\affiliation{Russian Quantum Center, Skolkovo, Moscow 121205, Russia}

\author{Nikolay N.~Kolachevsky}
\affiliation{P.N. Lebedev Physical Institute of the Russian Academy of Sciences, Moscow 119991, Russia}
\affiliation{Russian Quantum Center, Skolkovo, Moscow 121205, Russia}

\begin{abstract}
The use of multilevel quantum information carriers, also known as qudits, attracts a significant deal of interest as a way for further scalability of quantum computing devices. 
However, a non-trivial task is to experimentally achieve a gain in the efficiency of realizing quantum algorithms with qudits since higher qudit levels typically have relatively short coherence times compared to qubit states.
Here we propose and experimentally demonstrate two approaches for the realization of continuous dynamical decoupling of magnetic-sensitive states with $m_F=\pm1$ for qudits encoded in optical transition of trapped $^{171}$Yb$^{+}$ ions.
We achieve improvement in qudit levels coherence time by the order of magnitude (more than 9 ms) without any magnetic shielding, which reveals the potential advantage of the symmetry of the $^{171}$Yb$^{+}$ ion energy structure for counteracting the magnetic field noise.
Our results are a step towards the realization of qudit-based algorithms using trapped ions.
\end{abstract}

\maketitle

\section{Introduction}

The development of quantum computers is stimulated by the idea of achieving significant speedups over machines based on classical principles in solving computational tasks, 
which are related to cryptography~\cite{Shor1994}, search~\cite{Grover1996}, optimization~\cite{Farhi2014}, simulation of quantum systems~\cite{Lloyd1996}, solving large systems of linear equations~\cite{Lloyd2009}, etc. 
Existing prototypes of quantum computing devices use various physical platforms,
such as superconducting circuits~\cite{Martinis2019,Pan2021-4}, semiconductor quantum dots~\cite{Vandersypen2022,Morello2022,Tarucha2022}, optical systems~\cite{Pan2020,Lavoie2022}, 
neutral atoms~\cite{Lukin2021,Browaeys2021,Browaeys2020-2,Saffman2022}, and trapped ions~\cite{Monroe2017,Blatt2012,Blatt2018}, to implement quantum computing protocols.
Although several experiments have reported on achieving quantum advantage in solving sampling problems~\cite{Martinis2019,Pan2021-4,Pan2020}, 
the computational power of existing generation of quantum computers is limited.
These limitations are related to the fact that in order to solve a practically-relevant computational problem, the scalability of the devices with respect to the number of used information carriers 
(for example, qubits, which are quantum counterparts of classical bits) is needed to be combined with the high degree of quality of operations under the information carriers.
The combination of the quantitive and qualitative characteristics of quantum computing devices is the underlying idea of quantum volume (QV)~\cite{Gambetta20192}, which is one of the metrics for quantum computing power.
Trapped ions are one the first systems proposed for quantum computing~\cite{CiracZoller1995,Wineland1995}, 
which today demonstrates the highest QV of $2^{15}=32768$~\cite{Quantinuum2022}.
The reasons behind these results are the ability to demonstrate high-fidelity operations between the qubits~\cite{Wineland2016} and long coherence times~\cite{Kim2021}, 
which also allows one to prototype quantum error correction techniques~\cite{Wineland2004,Blatt2011,Blatt2020,Monroe2021,Blatt2021-2,Postler2022,Ryan-Anderson2022}.
However, for trapped ions, as for other platforms for quantum computing, the problem of scaling the system to large number of qubits with preserving high-enough fidelity of quantum gates is an outstanding challenge~\cite{Sage2019,Fedorov2022}. 
In particular, for trapped-ion-based systems, the number of controlled information carriers within a single trap is about 20-30.
Various techniques are considered as a method for further increasing the number of controlled ions~\cite{Sage2019,Monroe2013}.

The structure of information carriers in the quantum domains is typically much more complex, and it is artificially restricted to fit the conventional binary paradigm.
In particular, the structure of the Hilbert space of trapped ions~\cite{Senko2020,Ringbauer2021,Semerikov2022}, neutral atoms~\cite{Weggemans2022}, photonic systems~\cite{White2009,Morandotti2017,Chi2022}, 
and superconducting circuits~\cite{Martinis2009,Wallraff2012,Gustavsson2015,Martinis2014,Ustinov2015,Hill2021} that are used in quantum information processing,
admits much more complex quantum information encoding (for a review, see Ref.~\cite{Sanders2020}), 
which is the essence of qudit-based quantum computing~\cite{Farhi1998,Kessel1999,Kessel2000,Kessel2002,Muthukrishnan2000,Nielsen2002,Berry2002,Klimov2003,Bagan2003,Vlasov2003,Clark2004,Leary2006,Ralph2007,White2008,Ionicioiu2009,Ivanov2012,Li2013,Kiktenko2015,Kiktenko2015-2, Song2016,Frydryszak2017,Bocharov2017,Gokhale2019,Pan2019,Low2020,Jin2021,Martinis2009,White2009,Wallraff2012,Mischuck2012,Gustavsson2015,Martinis2014,Ustinov2015, Morandotti2017,Balestro2017,Low2020,Sawant2020,Senko2020,Pavlidis2021,Rambow2021,OBrien2022,Nikolaeva2022}. 
During last years, mutliqudit quantum processors~\cite{Hill2021,Ringbauer2021,OBrien2022,Semerikov2022}, 
including systems based on $^{40}$Ca$^{+}$~\cite{Ringbauer2021} and $^{171}$Yb$^{+}$~\cite{Semerikov2022} qudits, have been demonstrated. 
The use of qudits potentially allows one to realize quantum algorithms in a more efficient way~\cite{Nikolaeva2021}.
This is possible, first, since several qubits can be encoded into a one qudit~\cite{Kessel1999,Kessel2000,Kessel2002,Kiktenko2015,Kiktenko2015-2}; for example, a ququart processor is formally equivalent to a two-qubit processor, 
however, the quality of operations within a single trapped ion in the ququart case may be higher rather than for a two-qubit gate involving physical interaction between two ions.
Second, the use of additional qudit levels to substitute ancilla qubits in multiqubit gate decompositions~\cite{Barenco1995}
(for example, for the Toffoli gate~\cite{Ralph2007,White2009,Ionicioiu2009,Wallraff2012,Kwek2020,Baker2020,Kiktenko2020,Kwek2021,Galda2021,Gu2021}). 
It is interesting to mention that the first realization of two-qubit gates has used two qubits stored in a single trapped ion, i.e. in the qudit setup~\cite{Wineland1995}.
Thus, qudit-based quantum information processing can be considered as an useful paradigm for increasing the power of quantum computing devices.

The serious challenge is, however, to ensure long coherence times for the qudit states. 
For example, in the experimental demonstration of trapped-ion-based qudits on $^{40}$Ca$^{+}$~\cite{Ringbauer2021} with controlling up to seven levels, magnetic shielding is proposed as a solution to this problem. 
Our previous experimental setup has been based on optical qudits encoded in a quadrupole transition at 435 nm in $^{171}$Yb$^{+}$~\cite{Semerikov2022} 
with the use of magnetic-sensitive Zeeman sublevels of the upper level.
We note that this transition is typically used in quantum metrology~\cite{Tamm2000,Kersale2016,Leute2016,Kolachevsky2018}, 
and it has been used as for qubit encoding with ytterbium ions~\cite{ZalivakoSemerikov2021}. 
However, such a qudit structure has led to relatively short coherence times, which is the limitation for the realization of quantum algorithms~\cite{Semerikov2022}.

In this work, we demonstrate two approaches for the realization of continuous dynamical decoupling of magnetic-sensitive states with $m_F=\pm1$ for qudits encoded in $^{171}$Yb$^{+}$ trapped ions.
The simplicity of the presented techniques is due to the symmetry of the $^{171}$Yb$^{+}$ ion energy structure for counteracting the magnetic field noise and a large quadratic Zeeman shift lifting degeneracy 
of the transitions between neighboring Zeeman sublevels.
We achieve qudit levels coherence time of more than 9 ms without any magnetic shielding.
Our results indicate on potential advantages of ytterbium-ion-based qudits and open up the way to the realization of qudit-based algorithms.

Our work is organized as follows.
In Sec.~\ref{Sec:DD}, we describe the basics of continuous dynamical decoupling for ytterbium-based qudits.
We also present the method for radiofrequency dressing of $^{2}D_{3/2}$ manifold qudit states.
In Sec.~\ref{Sec:manipulation}, we describe the realization of operations with qudits.
In Sec.~\ref{Sec:reults}, we provide the results of the experimental implementation of two dynamical decoupling schemes.
We conclude in Sec.~\ref{sec:conclusion}.

\section{Continuous dynamical decoupling of $^{171}$Yb$^{+}$ optical qudit}\label{Sec:DD}

\subsection{Basic principles of continuous dynamical decoupling}

Dynamical decoupling is a well-known technique for suppressing decoherence~\cite{Lloyd1998,Lloyd1999,Knill2003}.
The detailed description and the comparison of different dynamical decoupling methods with the focus on trapped ions can be found, for example, in Ref.~\cite{Valahu2022}.
Here we only summarize basic principles of continuous dynamical decoupling (CDD) and multilevel continuous dynamical decoupling (MCDD) methods. 
Let us consider a pair of Zeeman components with opposite magnetic quantum numbers. 
For simplicity, we consider levels $|{-1}\rangle$ and $|1\rangle$ with $m_F=-1$ and $m_F=1$, respectively. 
The action of magnetic field fluctuations on this two-level system in the first order can be described by the following Hamiltonian:
\begin{equation}
	H_B=-g_F\mu_B\delta B\left(t\right)\sigma_z,
\end{equation}
where $g_F$ is the $g$-factor for the total angular momentum $F$, $\mu_B$ is the Bohr magneton, $\delta B\left(t\right)$ is the magnetic field fluctuations, and $\sigma_z$ is the Pauli matrix. 
Thus, as eigenenergies of these levels linearly depend on the magnetic field, their phases rapidly decohere in the noisy environment, which in turns limits their application in quantum information processing and prevents their use as qudit states. 

In order to suppress dephasing due to magnetic field fluctuations, we need to find or engineer states with energies which are weakly depending on the field magnitude. 
The part of the solution seems to store information in superposition states 
\begin{equation}
\begin{split}
	\left.|+\right\rangle=\frac{1}{\sqrt2}\left(\left.\left.|1\right\rangle+
    |-1\right\rangle\right), \\
	\left.|-\right\rangle=\frac{1}{\sqrt2}\left(\left.\left.|1\right\rangle-
    |-1\right\rangle\right),
\end{split}
\end{equation}
as due to symmetric influence of the Zeeman shift on both $|1\rangle$ and $|-1\rangle$ in their basis, $H_B$ takes form
\begin{equation}
	H_{B\pm}=-g_F\mu_B\delta B\left(t\right)(|+\rangle\langle-|+|-\rangle\langle-|).
\end{equation}
Here we can see, that the magnetic field noise does not change the mean energies of these states, so they do not dephase in the interaction picture. 
However, fluctuations in this basis couple $\left.|+\right\rangle$ and $\left.|-\right\rangle$ states, and cause the population transfer between them. This noise-induced population transfer has a resonant nature, namely mostly caused by the noise Fourier components at frequencies close to the energy difference between $|-\rangle$ and $|+\rangle$ states, which is zero in our case. Thus, constant field shifts and low-frequency noise components mostly contribute to this effect. This is especially inconvenient as in experiment low-frequency noise and particularly harmonics of mains power supply frequency (50 Hz) are dominating the noise spectrum. 

The idea behind the CDD approach is to lift these states degeneracy by application of dressing electromagnetic field, coupling levels $\left.|1\right\rangle$ and $\left.|-1\right\rangle$. 
States $\left.|+\right\rangle$ and $\left.|-\right\rangle$ diagonalize the resulting Hamiltonian, but this time their eigenenergies differ by a Rabi frequency $\Omega$ of the dressing field. 
In this case low-frequency noise is not be able to cause the population transfer between these states anymore as it is off-resonant.  
More precisely, to excite transitions between $\left.|+\right\rangle$ and $\left.|-\right\rangle$ noise frequency should be close to $\Omega$, or its magnitude must be large enough to broaden this transition. 
These conditions can be written as follows:
\begin{equation}
	\frac{\left|g_F\mu_B\widetilde{\delta B}\left(\omega\right)\right|}{\hbar\left|\omega-\Omega\right|}\geq1,
\end{equation}
where $\widetilde{\delta B}\left(\omega\right)$ is the Fourier amplitude of the noise. 
Thus, if dressing field Rabi frequency $\Omega$ is larger than the bandwidth of the noise and than the Larmor precession frequency of the state under influence of the noise, then the noise effect on the states is suppressed. 
This technique is known as the CDD approach. 
Although decoupling from magnetic field fluctuations monotonically improves with increase of $\Omega$, other factors cause degradation of the method’s performance if $\Omega$ is too large, which determines existence of an optimal value. 
These factors include increasing dephasing due to the dressing field amplitude fluctuations and off-resonant coupling to other states. 

We note that there are more sophisticated techniques to decouple these sources of errors too, such as usage of concatenated \cite{Cai_2012} or modulated \cite{Farfurnik2017} dressing fields, however they are beyond the scope of the present work.
Above we considered only a two-level system, however this method can be generalized to multiple states becoming MCDD. 

\subsection{Radiofrequency dressing of $^{2}D_{3/2}$ \\ manifold qudit states}

\begin{figure}
\center{\includegraphics[width=1\linewidth]{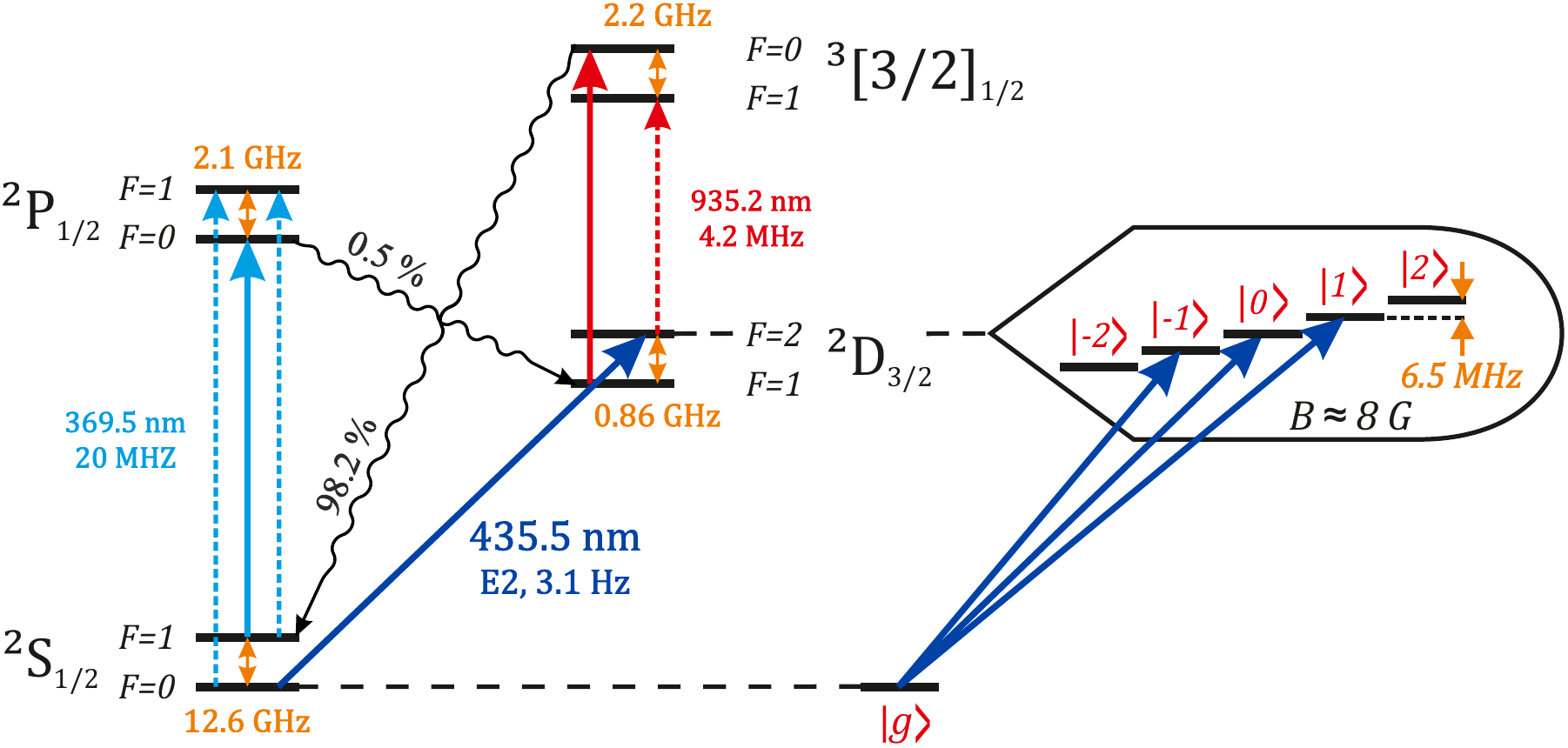}}
	\vskip-3mm
\caption{$^{171}$Yb$^{+}$ level structure used for qudit manipulations. Solid lines show laser fields used for laser cooling, state initialization, manipulation and readout. 
Dotted lines correspond to modulation sidebands obtained with electro-optical modulators and preventing population trapping in metastable hyperfine sublevels.}
\label{fig:levels}
\end{figure}

In this paper, we focus on decoupling $^{2}{D}_{3/2}$ ($F=2$, $m_F=\pm1$) states ($|{-1}\rangle$ and $|1\rangle$, respectively) of the $^{2}D_{3/2}$ manifold from the magnetic noise using radiofrequency fields (see Fig.~\ref{fig:levels}). 
These states along with $|0\rangle=\,^{2}{D}_{3/2}$ ($F=2$, $m_F=0$) and $|g\rangle=\,^{2}{S}_{1/2}$ ($F = 0$, $m_F = 0$) form a ququart, which is a qudit with four states for information encoding. 
We note that two other Zeeman sublevels with $m_F=\pm2$ can be also employed for information encoding, however in this paper we do not use them for this purpose. 
This is due to the fact that from the quantum information processing prospective, an ion string consisted of $N$ 4-level systems can be easily mapped and interpreted as $2N$ qubit processor. 
Also, usage of these states would require more sophisticated decoupling schemes to protect them from magnetic field fluctuations. 
We note, however, that we still tracked their populations in our experiments to detect leakages from the qudit subspace during experiments. 

As it follows from the previous subsection,  to protect a pair of magnetic sublevels from noise it is required to couple them with electromagnetic field. 
However, the selection rules prevent direct coupling of $|{-1}\rangle$ and $|1\rangle$ states with magnetic-dipole radiofrequency field, so a $|0\rangle$ state is used for that as a mediator. 

\begin{figure}
\centering
	\subfloat[Monochromatic radiofrequency dressing]
	{\includegraphics[width=1\linewidth]{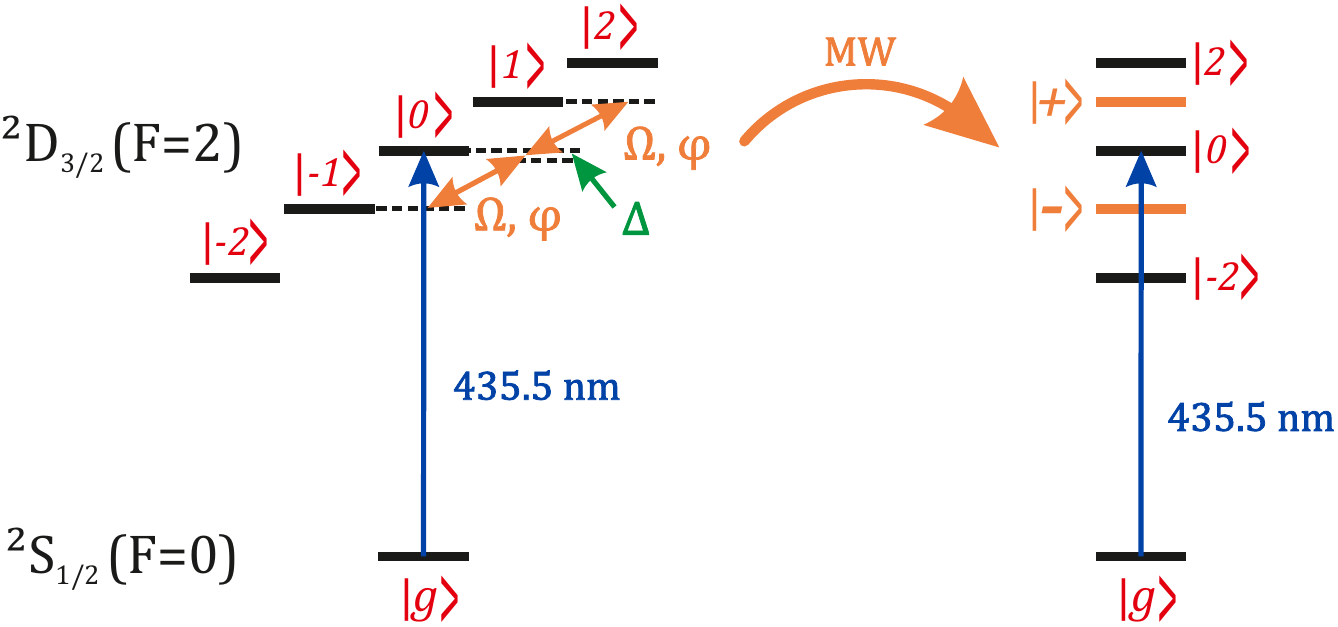}\label{fig:DDa}} \\ 
	\subfloat[Bichromatic radiofrequency dressing] 
	{\includegraphics[width=1\linewidth]{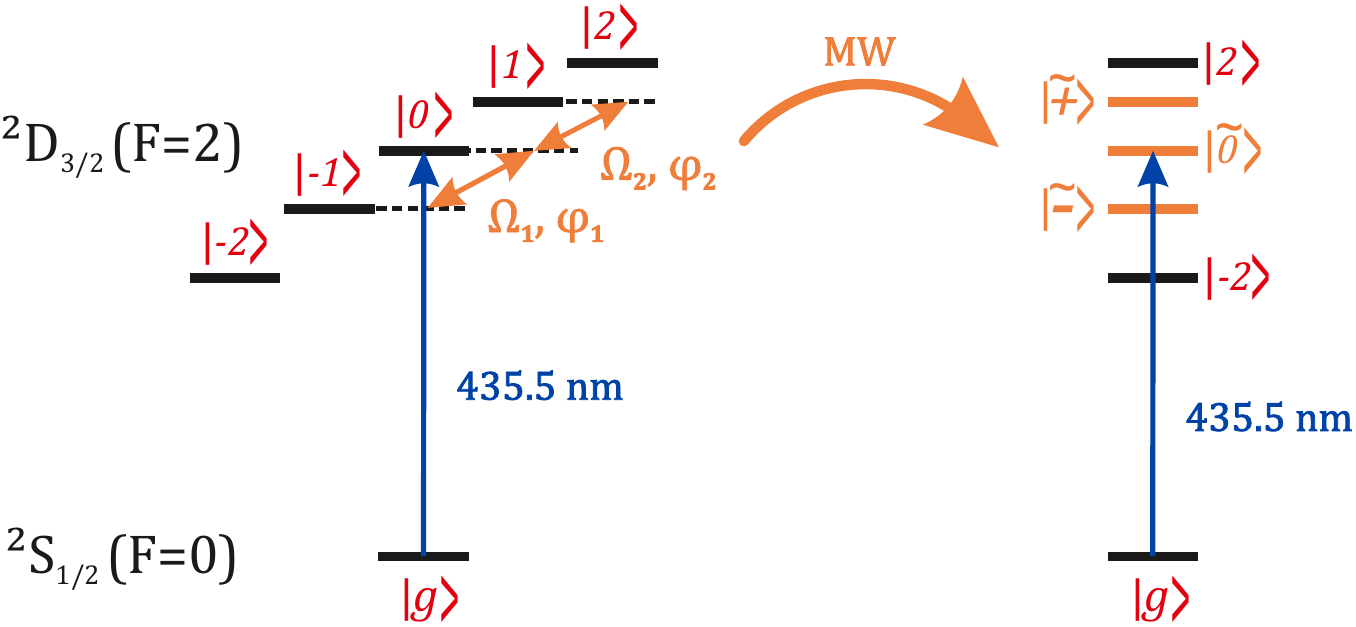}\label{fig:DDb}}
\caption{Dynamical decoupling schemes under consideration. 
Orange arrows on the left side of the pictures show decoupling fields with corresponding Rabi frequencies and phases. 
Right side shows eigenstates of the dressed ion. 
Black lines show states unaffected by the dressing, while orange correspond to superposition eigenstates determined by the interaction with the dressing fields.}
\end{figure}

Let us the consider a 3-level system of states $|{-1}\rangle$, $|{0}\rangle$, $|{1}\rangle$ with the corresponding Hamiltonian of the following form:
\begin{equation}
	H_0=\hbar\left(\begin{matrix}\omega_{-1}&0&0\\0&0&0\\0&0&\omega_1\\\end{matrix}\right).
\end{equation}
We note that due to quadratic Zeeman shift $\left|\omega_{-1}\right|\neq{|\omega}_1|$. 
We then add dressing radiofrequency fields with Rabi frequencies, field frequencies and phases ${\Omega_1,\omega_{D1},\phi_1}$ and ${\Omega_2,\omega_{D2},\phi_2}$, coupling $|-1\rangle$ with $|0\rangle$ and $|0\rangle$ with $|1\rangle$, respectively. 
We also place the constraint on fields frequencies, such that $\omega_{D1}+\omega_{D2}=\omega_1-\omega_{-1}$. It means that a sum frequency of dressing fields is equal to the spacing between $|-1\rangle$ and $|1\rangle$ states.
That enables us to parametrize frequencies by a single detuning $\Delta=\omega_{D2}-\omega_1$ (see Fig. \ref{fig:DDa}). 
Therefore, our consideration reduces to the classical problem of a $V$-system. 
The interaction with dressing fields will be described by the following Hamiltonian: 
\begin{equation}
\begin{split}
	H_D&=\frac{\hbar\Omega_1}{2}e^{-i\omega_{D1}t-i\phi_1}\left.|0\right\rangle\left\langle-1|\right.\\
	&+\frac{\hbar\Omega_2}{2}e^{-i\omega_{D2}t-i\phi_2}\left.|1\right\rangle\left\langle0|\right.+h.c.\\
	&=\frac{\hbar\Omega_1}{2}e^{i{(\omega}_{-1}+\Delta)t-i\phi_1}\left.|0\right\rangle\left\langle-1|\right.\\
	&+\frac{\hbar\Omega_2}{2}e^{-i(\omega_1+\Delta)t-i\phi_2}\left.|1\right\rangle\left\langle0|\right.+h.c.
\end{split}
\end{equation}
Moving to the rotation frame that is determined by the dressing fields and applying the rotating wave approximation gives us the total system Hamiltonian of the form:
\begin{equation}
	H_R=\frac{\hbar}{2}\left(\begin{matrix}-2\Delta&\Omega_1e^{i\phi_1}&0\\{\Omega_1e}^{-i\phi_1}&0&\Omega_2e^{i\phi_2}\\0&\Omega_2e^{-i\phi_2}&-2\Delta\\\end{matrix}\right).
\end{equation}
Exact expressions for eigenstates and eigenvalues of this Hamiltonian are rather bulky, so we do not provide them here explicitly. 
Two particular cases are considered below in detail.

\section{Decoupling schemes}

\subsection{Monochromatic decoupling scheme}

In the first scheme, we consider the case with the use of a single dressing field, which couples states $|-1\rangle$ and $|1\rangle$ via Raman two-photon transition.  
In this case $\Delta=-\frac{\omega_1+\omega_{-1}}{2}$, $\phi_1=\phi_2=\phi$, and $\Omega_1=\Omega_2=\Omega\ll\Delta$. 
This case is illustrated in Fig.~\ref{fig:DDa}. 
The scheme is equivalent to the classical two-level CDD scheme, but the two-photon process helps to avoid selection rules problem. 
If we put $\phi=0$ (only relative phases in the system make any difference, so this does not reduce generality), 
eigenstates for this case are $\{|0\rangle, |+\rangle, |-\rangle\}$ and their eigenenergies are $E=\{{\Omega^2}/{(2\Delta)},-\Delta-{\Omega^2}/{(2\Delta)},-\Delta\}$, respectively. 
Here one can note that the degeneracy of $|+\rangle$ and $|-\rangle$ states is lifted by the value of the effective Raman Rabi frequency $\Omega_e=\frac{\Omega^2}{2\Delta}$.
 At the same time $|0\rangle$ state experiences the Stark shift by the same value.
 
Such a scheme has its own advantages and disadvantages. 
As it has been discussed in the previous section, lifting the degeneracy of $|+\rangle$ and $|-\rangle$ states allows protecting them from the magnetic field noise with frequency and effective amplitude below $\Omega_e$. 
At the same time this scheme affects only states $|-1\rangle$ and $|1\rangle$ and leaves all others unperturbed (except the Stark shift on the $|0\rangle$).
Thus, all quantum control techniques involving other states can be used without any changes. 
For example, a two-qudit M$\o$lmer-S$\o$rensen gate (MS)~\cite{Blatt2003-2,Molmer-Sorensen1999,Molmer-Sorensen1999-2,Molmer-Sorensen2000} gate with states $|g\rangle$ and $|0\rangle$ can be performed exactly as it takes place for the qubit case. 
Also, as it is shown below, such qudit states can be rapidly and efficiently controlled with bichromatic laser fields, which is not a significant complication of the setup as bichromatic fields are also necessarily used for MS gates, 
so their support is routinely embedded in the experimental setups. 
On the other hand, the noise bandwidth, which can be suppressed by this method, is significantly limited by the difference in the transition frequencies between the Zeeman components, 
as 
\begin{equation}
	\Omega_e\ll\Omega\ll\Delta=-\frac{\omega_1+\omega_{-1}}{2}.
 \end{equation}
Thus, for example, for the value of the magnetic field $B\approx8$ G only noise, whose value is well below approximately 1 kHz, can be suppressed. 
We would like to note that in low-noise or high-bias-field applications this method appears to be the most convenient. 

\subsection{Bichromatic decoupling scheme}

The scheme considered above is based on the fact that the radiofrequency field is relatively far-detuned from $\left|-1\right\rangle\rightarrow|0\rangle$ and $\left|0\right\rangle\rightarrow|1\rangle$ transitions, 
and, therefore, the population in $|0\rangle$ decouples from $\left|-1\right\rangle$ and $|1\rangle$ states. 
Alternatively, one can use a pair of radiofrequency fields resonant to $\left|-1\right\rangle\rightarrow|0\rangle$ and $\left|0\right\rangle\rightarrow|1\rangle$ transitions (see Fig.~\ref{fig:DDb}). 
This corresponds to the case of $\Delta=0$. 
Let us also assume that $\Omega_1=\Omega_2=\Omega$, $\phi_1=0$, and $\phi_2=\phi$ (as it is already mentioned, only relative phases are relevant). 
Eigenstates for such system are as follows:
\begin{equation}
\begin{split}
	&\left|\widetilde{+}\right\rangle=\frac{1}{2}\left(e^{i\phi}\ \middle|-1\right\rangle+\left|1\right\rangle+\sqrt2e^{i\phi}\ \left|0\right\rangle), \\ 
	&\left|\widetilde{-}\right\rangle=\frac{1}{2}\left(e^{i\phi}\ \middle|-1\right\rangle+\left|1\right\rangle-\sqrt2e^{i\phi}\ \left|0\right\rangle), \\
	&\left|\widetilde{0}\right\rangle=\frac{1}{\sqrt2}\left(-e^{i\phi}\ \middle|-1\right\rangle+\left|1\right\rangle).
\end{split}
\end{equation}
Corresponding eigenenergies are $E=\{\frac{\Omega}{\sqrt2},-\frac{\Omega}{\sqrt2},0\}$. 
This is a particular case of the MCDD approach. 
 
It is important to note here that state $\left|\widetilde{0}\right\rangle$ eigenenergy does not depend on either magnetic field magnitude, nor on the dressing Rabi frequency $\Omega$ and,
 therefore, it is protected from the fluctuations of both of these values. 
In turn, states $\left|\widetilde{+}\right\rangle$ and $\left|\widetilde{-}\right\rangle$ are protected from magnetic field fluctuations, but they are still susceptible to Rabi frequency noise. 

This scheme has an advantage of higher decoupling Rabi frequency, as here only condition $\Omega\ll-\left(\omega_1+\omega_{-1}\right)$ must be satisfied, which is more relaxed than for a monochromatic case. 
Thus, faster and larger magnetic field fluctuations can be suppressed. 
On the other hand, this encoding scheme resonantly involves state $|0\rangle$, which can no longer be used for information storage independently from $\left|-1\right\rangle$ and $\left|1\right\rangle$. 

\section{Qudits manipulation}\label{Sec:manipulation}
To use qudits for quantum computing not only requires a sufficiently long coherence time, but also allow one to implement fast and efficient control for single- and two-qudit operations. 
Here we discuss how gates can be performed with dressed qudits described above.

\subsection{Single-qudit operations}

\begin{figure}[t!]
\centering
\subfloat[Monochromatic CDD scheme]{\includegraphics[width=0.7\linewidth]{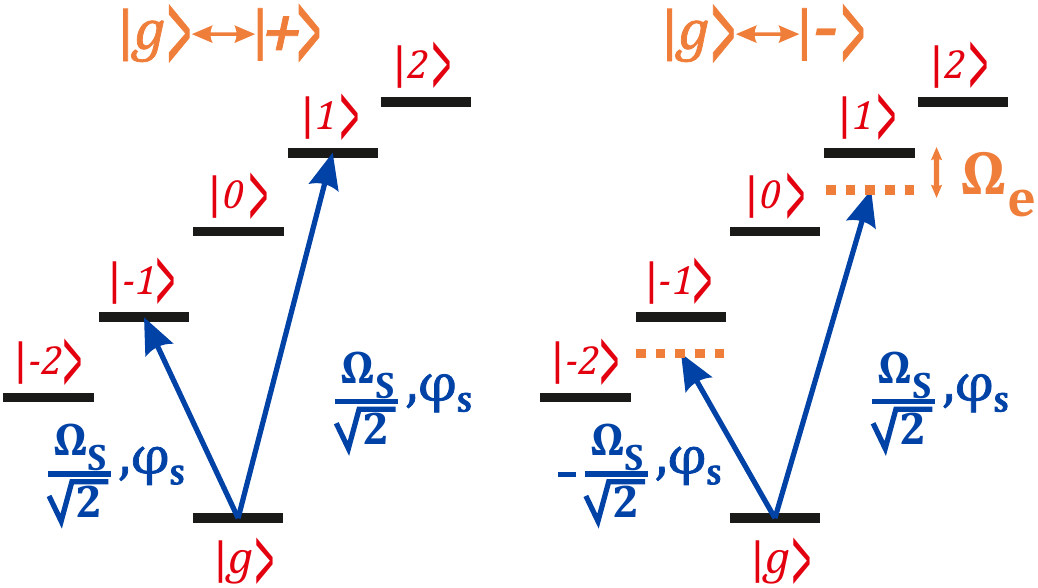}\label{fig:Addr_a}} \\
\subfloat[Bichromatic MCDD scheme]{\includegraphics[width=1\linewidth]{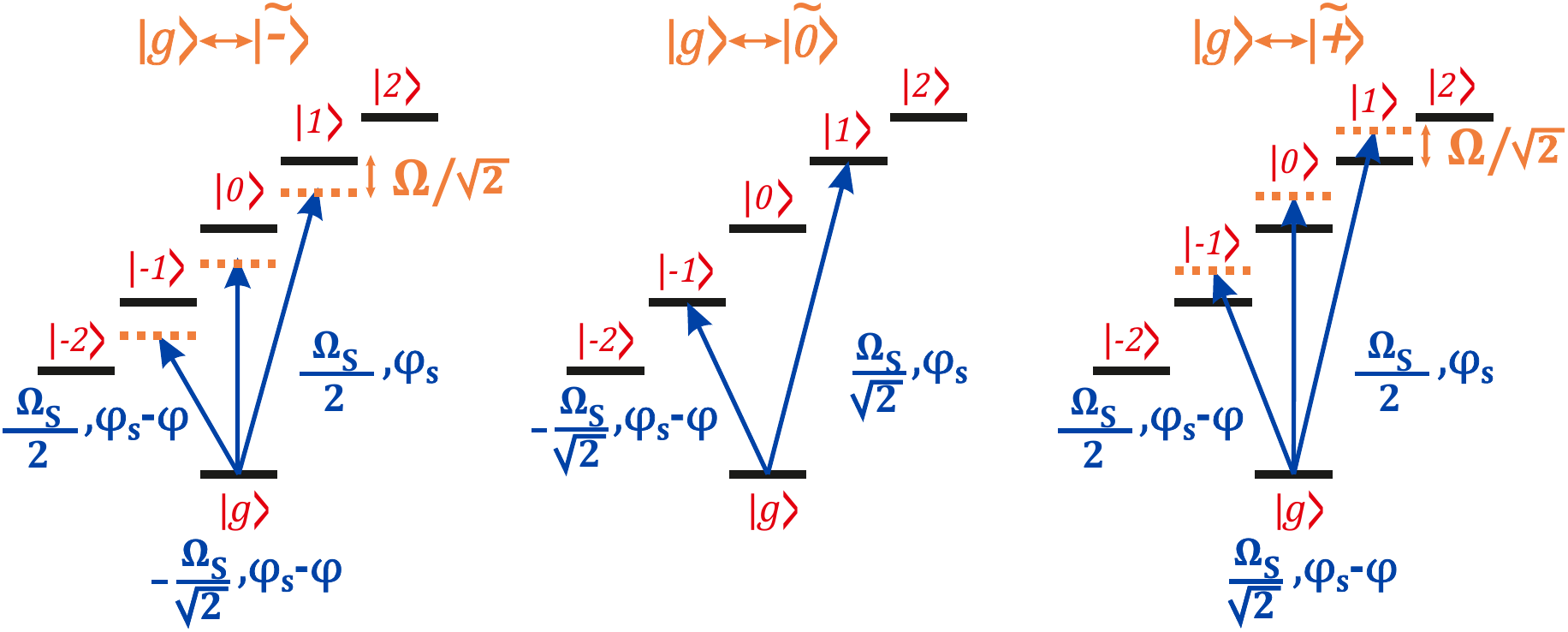}\label{fig:Addr_b}}
\caption{Configurations of the laser fields required to perform single-qudit operations for both decoupling schemes.}
\end{figure}

In case of conventional ion-based qubit setup, single-qubit operations are usually performed by application of a resonant electromagnetic field to the qubit transition which causes 
a state vector rotation on the Bloch sphere around some axis in the equatorial plane~\cite{Haeffner2008}. 
Rotation axis is determined by the relative field phase $\phi_S$ with respect to the qubit phase. 
The Hamiltonian generating such an operation in the interaction picture is given by
\begin{equation}
	H_{sq} = \frac{\hbar\Omega_S}{2}(e^{-i\phi_S}\sigma^+ + e^{i\phi_S}\sigma^-).
\end{equation}
Here $\Omega_S$ is the Rabi frequency of the field, $\phi_S$ is the field phase, determining rotation axis, and $\sigma^\pm$ are Pauli operators.

With undressed qudits we could use the same approach connecting state $|g\rangle$ with all other qudit state with a laser beam. 
As it can be shown explicitly (see Refs.~\cite{Semerikov2022, Blatt2021-3}), such operations constitute a universal single-qudit gate set. 
Using that fact we can derive control fields configurations required to generate analogous single-qudit gate set for dressed qudits case. 
In order to do so, we write down a required gate Hamiltonian in the dressed basis and interaction picture. 
Transforming it into original basis of Zeeman sublevels gives us required field configuration.
We note that transferring back to bare basis and its interaction picture causes appearance of oscillating terms, reflecting dressed states energy shifts.

For monochromatic case we have:
\begin{equation}
\begin{split}
	H_{sq}^{g+} &= \frac{\hbar\Omega_S}{2}(e^{-i\phi_S}|+\rangle\langle g| + h.c.) \rightarrow \\
	&\frac{\hbar\Omega_S}{2\sqrt{2}}\left[e^{-i\phi_S}(|1\rangle + |-1\rangle)\langle g| + h.c.\right],
\end{split}
\end{equation}
\begin{equation}
\begin{split}
	H_{sq}^{g-}&= \frac{\hbar\Omega_S}{2}(e^{-i\phi_S}|-\rangle\langle g| + h.c.) \rightarrow \\
	&\frac{\hbar\Omega_S}{2\sqrt{2}}\left[e^{-i\phi_S-i\Omega_et}(|1\rangle - |-1\rangle)\langle g| + h.c.\right].
\end{split}
\end{equation}

Therefore, corresponding single-qudit operations for monochromatically dressed qudits can be carried out with a bichromatic laser field (Fig. \ref{fig:Addr_a}). 
Its components are resonant (in case of $|g\rangle\rightarrow|+\rangle$ transition) or detuned by $-\Omega_e$ (for $|g\rangle\rightarrow|-\rangle$) from transitions between $|g\rangle$ and states $|-1\rangle$ and $|1\rangle$. 
The asymmetry between these states is caused by the Stark shift. 
Relative phase between Fourier components in this beam determine, which of the single-qudit operations to be performed (between $|g\rangle$ and $|+\rangle$ or $|g\rangle$ and $|-\rangle$). 
Similarly to the qubit case, the common phase $\phi_S$ of the beam components determines axis of the state vector rotation, and effective Rabi frequency of such an operation is by $\sqrt{2}$ larger than a Rabi frequency of each frequency component.  

We note that using this technique, we can engineer a Hamiltonian, which couples $|g\rangle$ with only one particular excited qudit state and, 
thus, duration of the operation is not limited by the necessity of spectroscopically resolving each of the $|g\rangle\to|\pm\rangle$ transitions (see Refs.~\cite{Webster2013,lahuerta2023}). 
At the same time, from the experimental point of view bichromatic addressing fields do not introduce significant additional complications as their support is anyway required for implementation of two-qubit gates, such as MS gate.

Similarly we can obtain field configuration for a case of bichromatic dressing, where required fields become more complicated:
\begin{equation}
\begin{split}
	&H_{sq}^{g\widetilde{+}} = \frac{\hbar\Omega_S}{2}(e^{-i\phi_S}|\widetilde{+}\rangle\langle g| + h.c.) \rightarrow \\
	&\frac{\hbar\Omega_S}{4}\left[e^{-i\phi_S+i\Omega t/\sqrt{2}}(e^{i\phi}|-1\rangle + |1\rangle + \sqrt{2}e^{i\phi}|0\rangle)\langle g| + h.c.\right],
\end{split}
\end{equation}
\begin{equation}
\begin{split}
	&H_{sq}^{g\widetilde{-}} = \frac{\hbar\Omega_S}{2}(e^{-i\phi_S}|\widetilde{-}\rangle\langle g| + h.c.) \rightarrow \\
	&\frac{\hbar\Omega_S}{4}\left[e^{-i\phi_S-i\Omega t/\sqrt{2}}(e^{i\phi}|-1\rangle + |1\rangle - \sqrt{2}e^{i\phi}|0\rangle)\langle g| + h.c.\right],
\end{split}
\end{equation}
\begin{equation}
\begin{split}
	&H_{sq}^{g\widetilde{0}} = \frac{\hbar\Omega_S}{2}(e^{-i\phi_S}|\widetilde{0}\rangle\langle g| + h.c.) \rightarrow \\
	&\frac{\hbar\Omega_S}{2\sqrt{2}}\left[e^{-i\phi_S}(-e^{i\phi}|-1\rangle + |1\rangle)\langle g| + h.c.\right].
\end{split}
\end{equation}
Thus, to excite $|\widetilde{\pm}\rangle$ states one needs to use trichromatic laser beams, while for $|\widetilde{0}\rangle$ state a bichromatic field is sufficient (Fig. \ref{fig:Addr_b}).

\subsection{Two-qudit operations}

In the ququart case, a single two-qudit MS type gate on any transition is sufficient to finish a universal multiqudit gate set~\cite{Semerikov2022, Blatt2021-3}. 
Therefore, a monochromatic decoupling scheme require no changes in ions entanglement procedure as states $|g\rangle$ and $|0\rangle$ ideally suited for such gate are not coupled to any others by the dressing fields. 
For bichromatic decoupling this is not the case, so entanglement procedure must be accordingly modified.

The protocol for such a gate can be derived in the same way as for a single-qudit operations. 
As dressing involves only internal degrees of freedom of ions, and commutation relations of the operators $|\widetilde{0}\rangle\langle g|$, $|g\rangle\langle \widetilde{0}|$, $|g\rangle\langle g|$ , $|\widetilde{0}\rangle\langle \widetilde{0}|$ 
are the same as for $|0\rangle\langle g|$, $|g\rangle\langle 0|$, $|g\rangle\langle g|$ , $|0\rangle\langle 0|$, all results for the conventional MS gate dynamics stay the same if we replace $|0\rangle$ in all operators to $|\widetilde{0}\rangle$. 
Therefore, in this case to perform a MS gate we need not a bichromatic laser field as usual MS gate, 
but a field with four spectral components, which are slightly detuned from red and blue secular sidebands of $|g\rangle\to|-1\rangle$ and $|g\rangle\to|1\rangle$ transitions.
Such fields can be readily generated using arbitrary-waveform generators feeding signal to acousto-optical modulators.

\section{Experimental results}\label{Sec:reults}

\subsection{Setup}

Ytterbium ions were stored in a linear Paul trap with single-particle secular frequencies of $\left\{\omega_x,\omega_y,\omega_z\right\}=2\pi\times\left\{3.88,\ 3.95,\ 0.16\right\}$ MHz, where $z$ is directed along the trap axis. 
The trap is installed inside a vacuum chamber with residual pressure below $10^{-10}$ mbar. 
All experiments were performed with an ion chain of 5 ions, which allowed efficient and single-shot readout of all Zeeman sublevels of the $^{2}D_{3/2}$ ($F = 2$) manifold.
Three orthogonal pairs of coils provided a magnetic field of 7.7 G, directed orthogonally to the trap axis. 
At this field transition frequencies between Zeeman components of the $^{2}D_{3/2}$ manifold are as follows: $\omega_{m_i,m_i+1}\approx2\pi\times\left[6465-11m_i\right]$ kHz, 
where $m_i$ is a magnetic quantum number of the level. 
Thus, $\omega_{-1,0}-\omega_{0,1}\approx2\pi\times11$ kHz.
We do not use magnetic shielding in this setup except synchronizing experiments start with mains supply line signal. 

Each experiment began with the Doppler cooling procedure with duration of 6 ms. 
The cooling process was carried out with a laser beam at 369 nm which was red-detuned from the $^{2}S_{1/2}$ ($F=1$) $\to$ $^{2}P_{1/2}$ ($F=0)$ quasi-cyclic transition~\cite{Zalivako2020}.
The beam was aligned at an the angle to all principal axes of the trap, which provided the cooling along all three of them. 
Another cooling beam along the trap axis assisted ion recrystallization in the case of crystal melting. 
Population trapping in $^{2}S_{1/2}$ ($F=0$) was avoided by the beam phase-modulation at 14.7 GHz with an electro-optical modulator (EOM), 
while the population leakage to $^{2}D_{3/2}$ due to spontaneous decay from the $^{2}P_{1/2}$ state was circumvented with the repumping laser at 935 nm. 
The latter is also phase-modulated at 3.07 GHz to remove population from both hyperfine components. 
Coherent population trapping, which could arise due to smaller total angular momentum of the $^{2}P_{1/2}$ ($F=0$) level with respect to the $^{2}S_{1/2}$ ($F=1$) state, 
was prevented by applying rather strong bias magnetic field and by the optimization of the beams polarizations~\cite{Berkeland2002,Ejtemaee2019}. 
After the cooling procedure, ions were optically pumped to the $^{2}S_{1/2}$ ($F=0$, $m_F=0$) state by turning off EOM at 14.7 GHz and turning on another EOM in this beam at 2.1 GHz. 
The process takes 10 $\mu$s, which is followed by blocking of cooling and repumping beams.

Optical qudit manipulations are performed with laser pulses at 435.5 nm. 
The fundamental harmonic of the external-cavity diode laser at 871 nm is frequency stabilized to the high-finesse optical ULE cavity. 
Details on the laser stabilization subsystem can be found in Ref.~\cite{Zalivako2020}. 
Laser emission is then passed through the tapered amplifier and the bow-tie cavity second-harmonic generator. 
The beam at 435 nm is passed through an acousto-optical modulator (AOM) for frequency, amplitude and phase control and an acousto-optical deflector providing the spatial steering of the beam and, therefore, the choice of the ion to be controlled. 
Then the beam is focused on the ions with a system of telescopes and an in-vacuum high-aperture lens. 
The system provides individual ion addressing with a cross-talk on the level of 3-10\%, depending on the ion. 
Signals for AOM and AOD are derived from several phase-coherent direct digital synthesizers (DDS). 
To generate multichromatic laser beams, which are required to control decoupled qudits, three DDS channels are combined together before AOM. 
Radiofrequency dressing fields are applied to one of the trap compensation electrodes. 
They are generated by another pair of DDS channels combined together. 
After combining the dressing signal is power amplified. 
These DDS channels are also phase-coherent with ones that are used for AOM driving. 
Thus, the setup enables one to control relative phases and amplitudes of all signals.

After required quantum operations are performed on all 5 ions, qudit states populations are detected via electron shelving technique~\cite{Semenin2021}.
In order to implement such a procedure, cooling and repumper beams are turned on again, however EOM at 3.07 GHz in the repumping beam is disabled. 
That prevents population pumping from $^{2}D_{3/2}$ ($F=2$) manifold to the $^{2}S_{1/2}$ --- $^{2}P_{1/2}$ subspace. 
In case of an ion being projected into $|g\rangle$ state strong fluorescence signal occurs, which is collected with a high-aperture lens onto a multi-channel PMT detector. 
In all other qudit states the fluorescence is suppressed. 
If the number of registered photons from a particular ion appears to be above some threshold, the the qudit is considered to be in state $|g\rangle$ and in certain other state otherwise. 
To measure the populations of all qudit states at the end of the experiment, we apply an additional $\pi$-pulse resonant to the $|g\rangle$$\to$ $|k-3\rangle$ transition to each of the ions before the readout occurs, 
where $k=\{1,…,5\}$ is the number of the particular ion. 
Thus, the probability of finding $k$th ion fluorescing is equal to the population of state $|k-3\rangle$ in the end of the experiment. 
The population of the last state $|g\rangle$ can be determined by the normalization condition.

\subsection{Dressed states manipulation and coherence}

\begin{figure}
\subfloat[Monochromatic CDD scheme]{\includegraphics[width=0.9\linewidth]{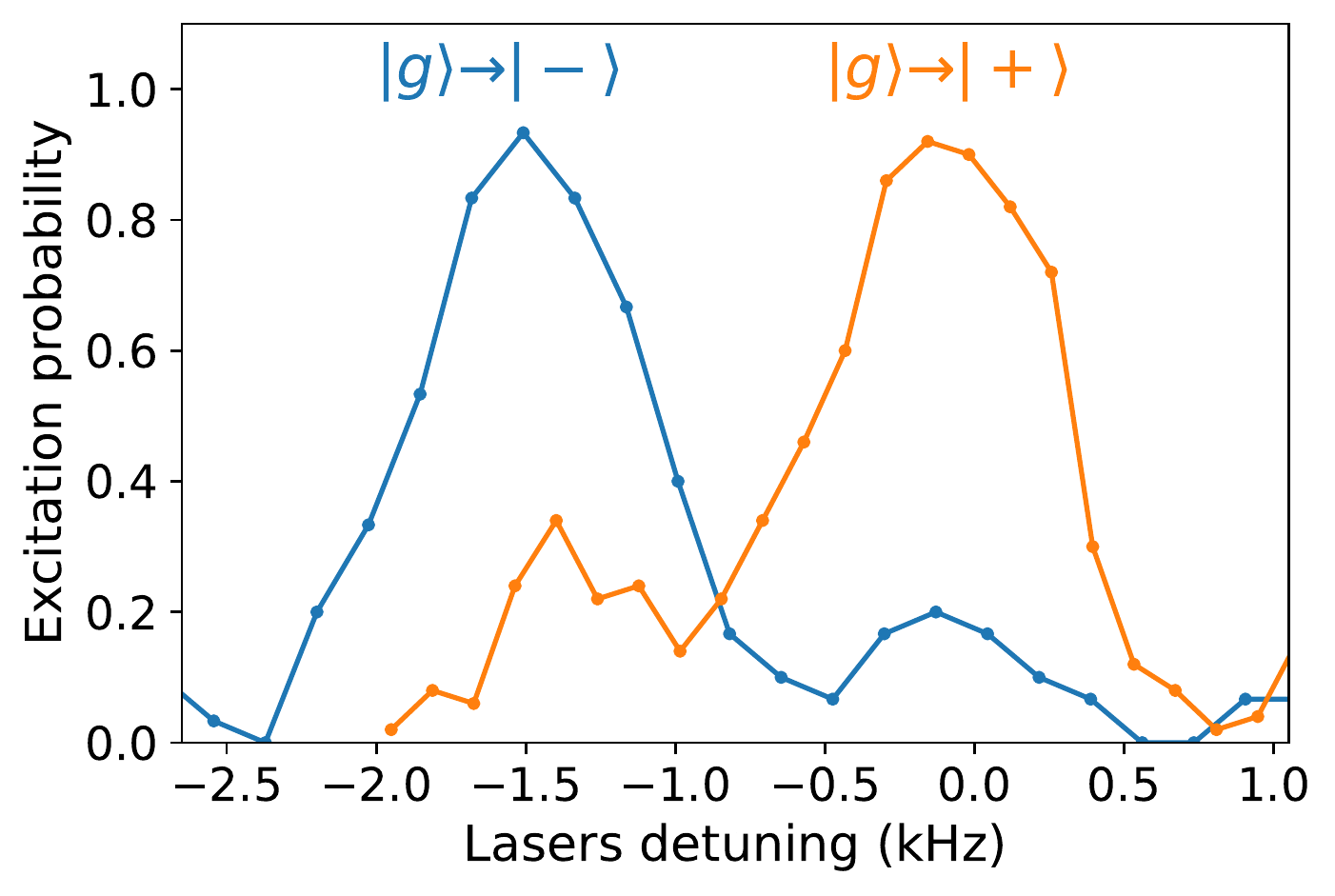}\label{fig:Spec_a}}
\\
\subfloat[Bichromatic MCDD scheme]{\includegraphics[width=0.9\linewidth]{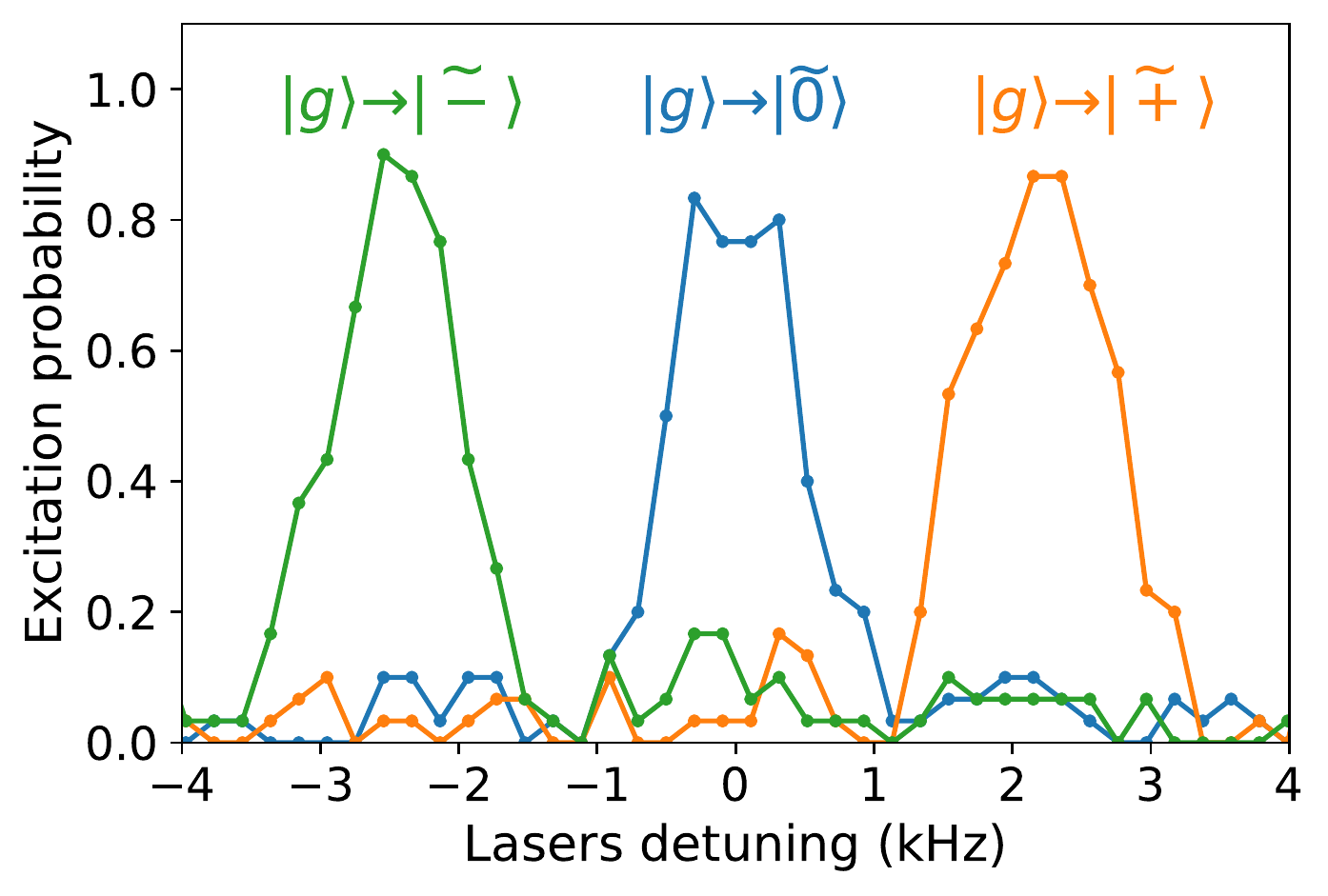}\label{fig:Spec_b}}
\caption{Spectroscopy of the dressed states after optimization of the addressing laser fields.
Different colors correspond to laser beams configurations engineered to address only a particular dressed state.}
\label{fig:Spectroscopy}
\end{figure}

Dressing fields frequencies and amplitudes were calibrated by observing Rabi-floppings between different Zeeman sublevels of the $\,^2D_{3/2}$ manifold. 
The effective Rabi frequency for monochromatic dressing was set to $\Omega_e=2\pi\times$1.5 kHz, and Rabi frequencies in bichromatic dressing to $\Omega_1=\Omega_2=\Omega=3.3$ kHz.

The relative phases and amplitudes of Fourier components in multichromatic addressing laser beams, 
which are required to interact with each particular dressed qudit state, were found spectroscopically to compensate for possible delays and losses in cables and signal combiners. 
To do so, parameters are initially set in according to theory given in the previous section. 
Then spectroscopy of all transitions between $|g\rangle$ and upper dressed qudit states was performed using weak and long laser pulses to resolve all transitions. 
Afterwards the relative amplitudes and phases were optimized to suppress excitation of any other transitions except the targeted one. 
Spectroscopy results after addressing fields optimization are shown in Fig. \ref{fig:Spectroscopy}.

In monochromatic case, one can see that parasitic dressed states excitations are not fully suppressed because of decoherence occurring due to magnetic field fluctuations. 
Dressing Rabi-frequency is evidently not sufficient in this case to protect the states from all fluctuations. Its increase, however, would cause too strong non-resonant excitation of $|0\rangle$ state. 
On the other hand, the bichromatic decoupling scheme enables one to apply stronger dressing fields and, thus, better decouple states from magnetic field fluctuations.

\begin{figure}
\center{\includegraphics[width=1\linewidth]{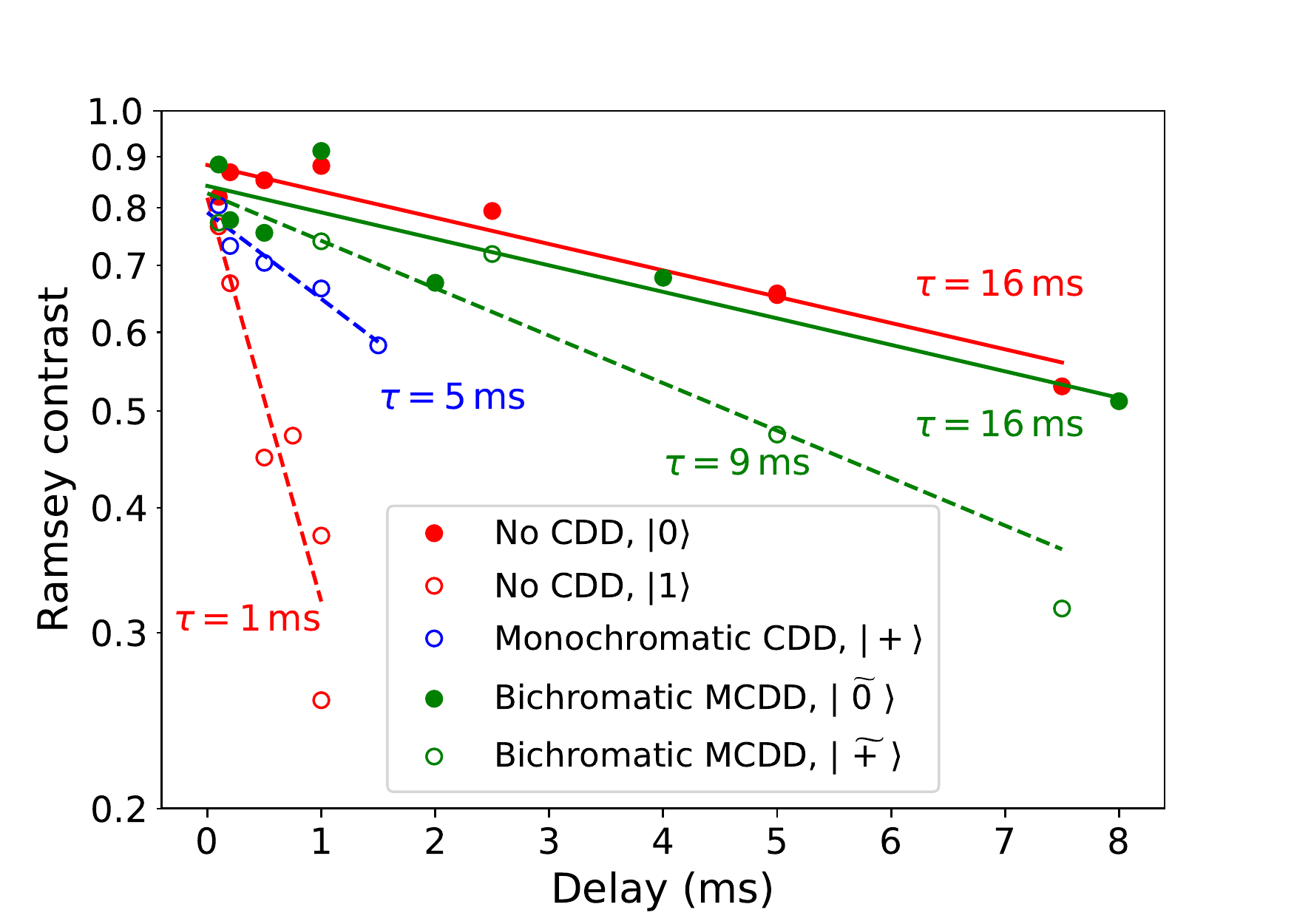}}
	\vskip-3mm
\caption{Dependence of the Raman fringes contrast on the delay between $\pi/2$ pulses. Different colors correspond to different decoupling schemes. 
For each scheme (except monochromatic) coherence time for several qudit states is measured. 
Data for fully decoupled states ($|0\rangle$ and $|\widetilde{0}\rangle$) is shown in filled circles, while unprotected ($|1\rangle$) or decoupling field amplitude sensitive states ($|+\rangle$ and $|\widetilde{+}\rangle$) are drawn with empty circles. 
One can see that without CDD coherence time for $|1\rangle$ is drastically shorter than for $|0\rangle$, while bichromatic MCDD ensures that all qudit states preserve coherence by an order of magnitude longer than the bare $|1\rangle$ state. 
The monochromatic CDD scheme shows intermediate results between bare states and bichromatic MCDD.}
\label{fig:Ramsey}
\end{figure}

The comparison of Ramsey contrast decay for different dynamical decoupling schemes and different states is presented in Fig.~\ref{fig:Ramsey}. 
In this measurement two $\pi/2$-pulses designed to excite the state under consideration were applied to the ion separated by a varying delay time. 
A common frequency shift was added to all addressing laser fields and its scanning gave rise to Ramsey fringes in excitation probability dependence (their contrast was measured). 
The dependence of the contrast on the delay between pulses was fitted with expression $f(t)=a\exp(-t/\tau)$, where $\tau$ is a coherence time ($T_2^*$) and $t$ is a delay between pulses.
The duration of the pulses is around 6 $\mu$s, which similar to ones used in quantum computing experiments for single-qubit operations, so it proves the feasibility of manipulation of dressed states manipulation at such rates.

As it can be seen from Fig. \ref{fig:Ramsey}, in the absence of dynamical decoupling, $|0\rangle$ state demonstrates relatively long coherence times of $T_2^{*}=16$ ms. 
This value is limited not by magnetic field fluctuations, but by the addressing laser frequency stability, which was confirmed by both laser stability comparison with other references and by repeating Ramsey measurements at different bias magnetic fields
affecting the level energy sensitivity to the fluctuations. 
The coherence time of magnetically sensitive $|1\rangle$ state, on the other hand, is only on the order of $1$ ms, which limits its applications for quantum computing. 

When monochromatic dynamical decoupling scheme is applied coherence time of dressed $|+\rangle$ and $-\rangle$ states is increased up to $5$ ms. 
However, as it is already mentioned above, the used efficient Rabi frequency is appeared to be not enough to suppress all noise, so this value is still lower than the one for bare $|0\rangle$. 
The bichromatic dressing scheme in this case demonstrates better results: dressed state $\left.|\widetilde{0}\right\rangle$ shows the same coherence time, as $0\rangle$, 
while $T_2^{*}$ for $\left.|\widetilde{+}\right\rangle$ and $\left.|\widetilde{-}\right\rangle$ reaches $9$ ms that is order of magnitude longer than initial values without dynamical decoupling.
Thus, our results indicate on the achieving qudit levels coherence time more than 9 ms without any magnetic shielding, 
which is two orders of magnitude longer than usual time for single-qudit operations and is sufficient for several dozens of two-qudit operations. 
This result can be further improved by application of stronger bias magnetic field and techniques to suppress dressing field fluctuations. 
Thus, our results reveals the potential advantage of the symmetry of the $^{171}$Yb$^{+}$ ion energy structure for counteracting the magnetic field noise.

\section{Discussion}\label{sec:discussion}

As it can be seen from the experimental results, both presented schemes have their advantages and disadvantages. 
The monochromatic CDD method is easier to implement in experiment as it does not require more than two spectral components in addressing beams and MS entangling operations also do not need any modifications with respect to undressed qudits or qubits. Thus, this method enables one to readily upgrade a conventional ytterbium optical qubit quantum processor to a qudit setup and realize qudit quantum algorithms. 
At the same time bichromatic scheme can suppress faster and larger magnetic field fluctuations which is clearly seen from the experimental data which results in longer coherence time. 
The price to pay for that is necessity to use control fields with three spectral components for single-qudit operations and four components for two-qudit gates, which still can be done with arbitrary waveform generators.  

As it is well known, calcium ion is another work-horse in the field of ion quantum computing and is also being actively studied as a platform for qudits implementation~\cite{Blatt2021-3}. 
Here we would like to point out that despite it is possible to achieve a CDD with this ion~\cite{lahuerta2023}, it is more complicated and less flexible than for ytterbium. 
Firstly, in case of ytterbium only dressing of the upper levels is required as $|g\rangle$ is intrinsically magnetically-insensitive. In case of calcium both upper and lower manifolds must be dressed. 
Secondly, as $\,^{40}Ca^+$ has a very small quadratic Zeeman shift and all transitions between neighboring sublevels are degenerate, radiofrequency dressing fields always mix all sublevels in the upper manifold. 
Both these factors significantly complicate the structure of the dressed states and manipulation process of such qudits. That inevitably increases contribution of the addressing errors to the gate fidelities. 
At the same time ytterbium, how it was shown in this paper, due to its symmetric energy structure and large quadratic Zeeman shift enables one to engineer simple and flexible decoupling schemes, 
involving only states of interest, and keep addressing system rather simple.

\section{Conclusion}\label{sec:conclusion}

Limitations that are related to short coherence times are of importance for the realization of quantum information processing.
In this work, we have addressed this problem by demonstrating two approaches for the realization of continuous dynamical decoupling of magnetic-sensitive states with $m_F=\pm1$ for qudits encoded in $^{171}$Yb$^{+}$ trapped ions.
As we have shown, qudit levels coherence time of more than 9 ms can be achieved without any magnetic shielding.
The key ingredient of the proposed schemes is the use of the symmetry of the $^{171}$Yb$^{+}$ ion energy structure for counteracting the magnetic field noise.
As we expect, our results are of importance for realizing qudit-based algorithms with trapped ions.

\section*{Acknowledgements}
We thank A.S. Nikolaeva and E.O. Kiktenko for fruitful discussions and useful comments.

\section*{Funding}
The experimental work supported by the Russian Roadmap on Quantum Computing (Contract No. 868-1.3-15/15-2021, October 5, 2021).

\bibliography{bibliography.bib}

\end{document}